\documentclass[11pt]{article}

\usepackage{amsfonts}
\usepackage{amsbsy} 
\usepackage{epsfig}
\usepackage{latexsym}
\input amssym.def
\input amssym.tex

\def\bequ{\begin{equation}}
\def\eequ{\end{equation}}
\def\barr{\begin{array}}
\def\earr{\end{array}}
\def\half{{1\over 2}}
\def\ben{\begin{equation}}
\def\een{\end{equation}}
\def\bena{\begin{eqnarray}}
\def\eena{\end{eqnarray}}

\advance \topmargin by -\headheight
\advance \topmargin by -\headsep
\evensidemargin \oddsidemargin
\advance\hoffset by -3mm  
\advance\voffset by  8mm  

\def\bR{\Bbb R}

\def\bH{\Bbb H}

\def\b1{e^0}

\begin{document}
\hfuzz=100pt
\title{{\LARGE \bf{The Petrov and Kaigorodov-Ozsv\'ath Solutions: Spacetime as a Group Manifold}}}
\author{Gary W. Gibbons\footnote{Electronic address: gwg1@damtp.cam.ac.uk},  Steffen Gielen\footnote{Electronic address: sg452@damtp.cam.ac.uk}
\\
\\ D.A.M.T.P.,
\\ Cambridge University,
\\ Wilberforce Road,
\\ Cambridge CB3 0WA,
\\ U.K.
}

\maketitle

\begin{abstract}
The Petrov solution (for $\Lambda=0$) and the Kaigorodov-Ozsv\'ath solution (for $\Lambda<0$) provide examples of vacuum solutions of the Einstein equations with simply-transitive isometry groups. We calculate the boundary stress-tensor for the Kaigorodov-Ozsv\'ath solution in the context of the adS/CFT correspondence. By giving a matrix representation of the Killing algebra of the Petrov solution, we determine left-invariant one-forms on the group. The algebra is shown to admit a two-parameter family of linear deformations a special case of which gives the algebra of the Kaigorodov-Ozsv\'ath solution. By applying the method of non-linear realisations to both algebras, we obtain a Lagrangian of Finsler type from the general first-order action in both cases. Interpreting the Petrov solution as the exterior solution of a rigidly rotating dust cylinder, we discuss the question of creation of CTCs by spinning up such a cylinder. We show geodesic completeness of the Petrov and Kaigorodov-Ozsv\'ath solutions and determine the behaviour of geodesics in these spacetimes. The holonomy groups were shown to be given by the Lorentz group in both cases.
\end{abstract}
\eject

\tableofcontents

\section{Introduction}
Among the wide range of known exact solutions to Einstein's field equations (the most comprehensive source is \cite{exact}), there are some with particularly interesting symmetry properties. In this paper we will concentrate on {\it homogeneous spacetimes}, which by definition admit a transitive group of motions; a group of motions (or {\it isometry group}) is a continuous (Lie) group of transformations preserving the metric of a given spacetime. Its generators are the Killing vector fields $\xi_A$ which satisfy
\bequ
(\mathcal{L}_{\xi_A}g)_{ab}=0\quad\Leftrightarrow\quad ({\xi_A})_{a;b}+({\xi_A})_{b;a}=0.
\eequ
That a group acts transitively means that the orbits under the group action are equal to the manifold itself, i.e. for any two given points $p$ and $q$ there is a transformation which maps $p$ to $q$. If this transformation is unique, the action is said to be simply transitive, otherwise it is multiply transitive; for a multiply transitive group action there exists for any given point a subgroup which leaves this point invariant, which is called the {\it stabilizer subgroup} (or little group) of that point. After choice of an ``origin" one may identify each spacetime point with all elements of $G$ that map the origin to this point. Stabilizer groups of different points are different but isomorphic, and hence one can identify the homogeneous space with the coset $G/K$, where $G$ is the group of motions and $K$ the stabilizer.
\\Elementary examples are provided by $n$-spheres with isometry group $SO(n+1)$ and stabilizer $SO(n)$,
\bequ
S^n \cong SO(n+1)/SO(n),
\eequ
and by Minkowski space ${\Bbb M}^n$
\bequ
{\Bbb M}^n \cong E(n-1,1)/SO(n-1,1),
\eequ
where $E(n-1,1)$ is the Poincar\'e group.
\\The existence of such a transitive group of motions allows one to use group theoretic methods to analyse the structure of the manifold.
\\
\\In the particularly simple case of a simply transitive group of motions, the stabilizers are trivial and hence there is a one-to-one correspondence between spacetime points and elements of the group of motions. Therefore the spacetime is not only a manifold but also a Lie group. One example, and in fact the only example among vacuum solutions without a cosmological constant, is provided by the Petrov solution which will be analysed in this paper.
\\
\\Our motivation to study homogeneous spacetimes comes from the method of {\it non-linear realisations}, where one looks for a general prescription to write down an action on a spacetime, interpreted as a coset $X=G/K$, that respects the given symmetry. It should be invariant under the action of a subgroup $H\subset K$ and transform in a ``covariant'' fashion under the action of $G$. One usually starts by considering left-invariant one-forms on the group $G$, then writes down an action from these and removes Lagrange multipliers from the action. We will give examples for this method when it is discussed in Section 3.
\\Since one only needs to specify the groups $G$ and $K$, the method promises to be very generally applicable and may be of interest in the development of theories beyond M-theory.
\\The general theory of non-linear realisations applied to internal symmetries was introduced in \cite{callan}. 
\\
\\We will present the Petrov solution with a possible physical interpretation as the exterior solution of an infinite rigidly rotating cylinder in Section 2, and also a related vacuum solution with negative cosmological constant which we will refer to as the Kaigorodov-Ozsv\'ath solution. For the latter we compute the stress-energy tensor of the boundary theory in the context of the adS/CFT correspondence from an expansion near the conformal boundary. The starting point of the analysis in Section 3 will be the isometry group of the Petrov spacetime and its Lie algebra. We will give a matrix representation of the group and the algebra and use it to construct left-invariant one-forms on the group. This gives the form of a general first-order action. Eliminating the non-dynamical variable from this action will give a Lagrangian of Finsler type, although we can not give an explicit form.
\\Since the isometry group is the basic object of interest here, we will then look for extensions of the Petrov solution by looking at {\it deformations} of the Lie algebra in Section 4. One of these deformations leads to the Killing algebra of the Kaigorodov-Ozsv\'ath solution. When discussing non-linear realisations for this case we will eliminate the non-dynamical variable and give an explicit form for the resulting Lagrangian which again is of Finsler type.
\\The Petrov solution contains closed timelike curves (CTCs), and in the context of possible causality violation by CTCs a central question is whether these can be created by some process in a spacetime which did not exhibit CTCs initially. Hawking formulated the chronology protection conjecture which asserts that the appearance of CTCs is forbidden by the laws of physics \cite{hawking92}. We will analyse the possible appearance of CTCs by spinning up a rotating cylinder in Section 5.
\\In a more detailed analysis of the physical properties of the Petrov and Kaigorodov-Ozsv\'ath solutions in Section 6, we will show that they are geodesically complete and we will obtain general statements about geodesics in these spacetimes. We calculate the holonomy groups of the spacetimes in Section 7. A few conclusions are given in Section 8.

\setcounter{equation}{0}
\section{Vacuum Solutions With Simply-Transitive Groups of Motions}

\subsection*{$\Lambda=0$ - The Petrov Solution}
The Petrov solution is introduced in \cite{exact} in the following theorem: {\it The only vacuum
solution of Einstein's equations admitting a simply-transitive
four-dimensional maximal group of motions is given by}
\bequ
k^2
ds^2=dr^2+e^{-2r}dz^2+e^r(\cos\sqrt{3}r(d\phi^2-dt^2)-2\sin\sqrt{3}r\,d\phi\,dt),
\label{petrov}
\eequ
where $k$ is an arbitrary constant, which shall be set equal to
one, and we have relabelled the coordinates compared to
\cite{exact}. The solution was first given in \cite{petrov} and also discussed in \cite{debever}. It describes a hyperbolic plane $\bH^2$ (the $(r,z)$-plane) with a timelike two-plane attached to each point.
\\The isometry group is generated by the Killing vector fields
\bequ
T\equiv\partial_t,\;\Phi\equiv\partial_{\phi},\;Z\equiv\partial_z,\;R\equiv\partial_r+z\partial_z+\half(\sqrt{3}t-\phi)\partial_{\phi}-\half(t+\sqrt{3}\phi)\partial_t,
\label{killing}
\eequ
which satisfy the algebra
\bequ
[R,T]=\half
T-\frac{\sqrt{3}}{2}\Phi,\;[R,\Phi]=\half\Phi+\frac{\sqrt{3}}{2}T,\;[R,Z]=-Z.
\label{algebra}
\eequ
The isometry group contains three-dimensional subgroups of Bianchi
Types $I$ and $VII_h$ acting in timelike hypersurfaces, and the
solution (\ref{petrov}) is Petrov type $I$ \cite{exact}. The first three Killing vectors obviously generate translations while the action of the one-parameter subgroup generated by $R$ on spacetime is given by the integral curves of $R$, which satisfy
\bequ
\frac{dx^a(\lambda)}{d\lambda}=R^a(x(\lambda)).
\eequ
These integral curves have the form
\bequ
x^a(\lambda)=\left(r_0+\lambda,z_0e^{\lambda},\phi_0e^{-\frac{\lambda}{2}}\cos\frac{\sqrt{3}}{2}\lambda+t_0e^{-\frac{\lambda}{2}}\sin\frac{\sqrt{3}}{2}\lambda,-\phi_0e^{-\frac{\lambda}{2}}\sin\frac{\sqrt{3}}{2}\lambda+t_0e^{-\frac{\lambda}{2}}\cos\frac{\sqrt{3}}{2}\lambda\right),
\eequ
where we label the coordinates by $x^a=(r,z,\phi,t)$.
\\
\\The metric components $g_{\phi\phi}$ and $g_{tt}$ become zero at certain values\footnote{these values of $r$ can be shifted by a coordinate transformation $\phi\rightarrow\alpha\phi+\beta t,\;t\rightarrow-\beta \phi+\alpha t$ and hence have no coordinate-independent significance} of $r$, but as the determinant of the metric in (\ref{petrov}) is always $-1$, it
is possible to extend the coordinates to infinite ranges and the coordinates ($r,z,\phi,t$) define a global chart. The manifold is also time-orientable, as the vector field $t^a=(0,0,\sin\frac{\sqrt{3}}{2}r,\cos\frac{\sqrt{3}}{2}r)$ defines a global arrow of time, though this amounts to $t$ being future- as well as past-directed at some points.
\\
\\Bonnor \cite{bonnor79} pointed out that the solution can be viewed as a special case of the exterior part of a Lanczos-van Stockum solution \cite{lanczos, vanstockum} describing an infinite cylinder of rigidly rotating dust. Since this allows a physical interpretation of (\ref{petrov}), let us give the general solution for
an infinite rigidly rotating dust cylinder, which, in
Weyl-Papapetrou form, is given by \cite{tipler74}
\bequ
ds^2=H(\rho)(d\rho^2+d\tilde{z}^2)+L(\rho)d\chi^2+2M(\rho)\,d\chi\,d\tau-F(\rho)d\tau^2,
\label{vanstockum}
\eequ
where $H,\;L,\;M,\;F$ are functions of the radial variable $\rho$
containing two parameters $a$ and $R$, interpreted as the angular
velocity and radius of the cylinder respectively.
\\The high energy case $aR>\frac{1}{2}$, which contains closed
timelike curves (CTCs), is of interest here, and with the choices
$R=\sqrt{e}$ and $aR=1$ the exterior solution is given by
\bequ
H(\rho)=\frac{1}{\rho^2},\;L(\rho)=-2\sqrt{\frac{e}{3}}\rho\sin\left(\sqrt{3}\log\frac{\rho}{\sqrt{e}}\right),\;M(\rho)=\frac{2}{\sqrt{3}}\rho\sin\left(\frac{\pi}{3}+\sqrt{3}\log\frac{\rho}{\sqrt{e}}\right),
\eequ
\[F(\rho)=\frac{2}{\sqrt{3e}}\rho\sin\left(\frac{\pi}{3}-\sqrt{3}\log\frac{\rho}{\sqrt{e}}\right),\]
where $\rho$ is a radial coordinate in the exterior of the cylinder and so is restricted to 
\bequ
\rho\ge\sqrt{e}, 
\label{rhobig}
\eequ
and $\chi$ is an angular coordinate and periodically identified with period $2\pi$; $\tau$ and $\tilde{z}$ are unconstrained. 
Applying the coordinate transformations
\[\tilde
z=\sqrt{e}\,z,\;\rho=\sqrt{e}\,e^r,\;\chi=\frac{1}{\sqrt[4]{3}\sqrt{2e}}\left(\sqrt{2+\sqrt{3}}\phi+\sqrt{2-\sqrt{3}}t\right),\tau=\frac{1}{\sqrt[4]{3}}(\phi-t)\]
to the line element (\ref{vanstockum}) indeed gives back (\ref{petrov}). Hence if we adopt the interpretation of (\ref{petrov}) as describing the exterior of a spinning cylinder, we restrict the coordinates to $r\ge 0$ and identify $\sqrt{2+\sqrt{3}}\phi+\sqrt{2-\sqrt{3}}t$ with $\sqrt{2+\sqrt{3}}\phi+\sqrt{2-\sqrt{3}}t+2\pi\sqrt[4]{3}\sqrt{2e}$.
\\
\\The general Lanczos-van Stockum solution has three linearly independent Killing vectors $\partial_{\chi},\;\partial_{\tau}$ and $\partial_z$ but a fourth Killing vector is only present in the special case $aR=1$ because the algebraic invariants of the Riemann tensor are independent of $\rho$ just in this case \cite{bonnor80}.
\\
\\Let us finally present all non-vanishing Christoffel symbols for the metric (\ref{petrov})
\bequ
{\Gamma^r}_{zz}=e^{-2r},\;{\Gamma^r}_{\phi t}=e^{r}\left(\frac{1}{2}\sin\sqrt{3}r+\frac{\sqrt{3}}{2}\cos\sqrt{3}r\right),\;{\Gamma^r}_{\phi\phi}=e^{r}\left(-\frac{1}{2}\cos\sqrt{3}r+\frac{\sqrt{3}}{2}\sin\sqrt{3}r\right),
\eequ
\[{\Gamma^r}_{tt}=e^{r}\left(\frac{1}{2}\cos\sqrt{3}r-\frac{\sqrt{3}}{2}\sin\sqrt{3}r\right),\;{\Gamma^z}_{rz}=-1,\;{\Gamma^{\phi}}_{r\phi}=\frac{1}{2},\;{\Gamma^{\phi}}_{rt}=-\frac{\sqrt{3}}{2},\;{\Gamma^t}_{r\phi}=\frac{\sqrt{3}}{2},\;{\Gamma^t}_{rt}=\frac{1}{2}.\]

\subsection*{$\Lambda<0$ - The Kaigorodov-Ozsv\'ath Solution}
A solution of the vacuum Einstein equations with {\it negative} cosmological constant which has a simply-transitive four-dimensional group of motions was first given by Kaigorodov \cite{kaigorodov} and rediscovered by Ozsv\'ath \cite{ozsvath}. It has the line element
\bequ
ds^2=-\frac{3}{\Lambda}dr^2+e^{-2r}(dz^2+2\,dt\,d\phi)+e^{4r}d\phi^2-2\sqrt{2}e^r\, dz\,d\phi.
\label{lambda}
\eequ
This solution was also given in \cite{exact} (with misprints, which were corrected in the second edition). It is Petrov type $III$ and the metric asymptotically (as $r\rightarrow -\infty$) approaches that of anti-de Sitter space. Obvious Killing vectors are 
\bequ
Z\equiv\partial_z,\;\Phi\equiv\partial_{\phi},\;T\equiv\partial_t,
\eequ
and the metric (\ref{lambda}) has a further isometry
\bequ
r\rightarrow r+\lambda,\;z\rightarrow e^{\lambda}z,\;\phi\rightarrow e^{-2\lambda}\phi,\;t\rightarrow e^{4\lambda}t
\eequ
which is generated by the fourth Killing vector $R\equiv\partial_r+z\partial_z-2\phi\partial_{\phi}+4 t\partial_t$. The Killing vector fields satisfy the algebra
\bequ
[R,Z]=-Z,\;[R,\Phi]=2\Phi,\;[R,T]=-4 T.
\label{algnew}
\eequ
No analogous solution for a positive cosmological constant exists \cite{exact, ozsvath}. Because of the similarity to (\ref{algebra}) one would expect (\ref{algnew}) to arise as a deformation of (\ref{algebra}). Physically both algebras describe the isometries of vacuum solutions of Einstein's equations, one with $\Lambda<0$ and one with $\Lambda=0$, and hence one might expect one of them to arise as some limit of the other.
\\This spacetime is also time-orientable, as the vector field $t^a=(0,-\frac{1}{\sqrt{2}}e^{3r},-1,1)$ defines a global arrow of time.
\\We can express (\ref{lambda}) in coordinates corresponding to Poincar\'e coordinates on adS (with $\Lambda=-3$)
\bequ
ds^2=\frac{d\rho^2+dz^2+d\phi^2-dt^2}{\rho^2}-2\rho\,dz(dt+d\phi)+\half\rho^4 (dt+d\phi)^2.
\label{poincare}
\eequ
The limit $\rho\rightarrow 0$ in Poincar\'e coordinates corresponds to the timelike boundary $\mathcal{I}$ of anti-de Sitter space. After setting $\tilde\rho=\rho^2$ the line element is
\bequ
ds^2=\frac{d\tilde{\rho}^2}{4\tilde{\rho}^2}+\frac{1}{\tilde\rho}\left(dz^2+d\phi^2-dt^2-2\tilde{\rho}^{3/2}\,dz(dt+d\phi)+\half\tilde{\rho}^3 (dt+d\phi)^2\right),
\label{poincare2}
\eequ
and (\ref{poincare2}) is an expansion of the form
\bequ
ds^2=\frac{d\tilde{\rho}^2}{4\tilde{\rho}^2}+\frac{1}{\tilde\rho}g_{ij}dx^i dx^j,\quad g_{ij}(x,\tilde{\rho})=g_{ij}^{(0)}(x)+g_{ij}^{(2)}(x)\tilde{\rho}+g_{ij}^{(3)}(x)\tilde{\rho}^{3/2}+\ldots
\eequ
as given in \cite{anomaly}. The coefficient $g_{ij}^{(3)}=-2(dz\otimes(dt+d\phi))_{ij}$ encodes the stress energy tensor of the boundary dual theory in the context of the adS/CFT correspondence \cite{anomaly, subra}. It does not satisfy even the null energy condition on the three-dimensional conformal boundary, since $g_{ij}^{(3)}n^i n^j=-2$ for the null vector $n^a=(\frac{\partial}{\partial t}+\frac{\partial}{\partial z})^a$. This presumably reflects causal pathologies of the bulk spacetime.
\\
\\A general analysis of stationary cylindrically symmetric Einstein spaces was done in \cite{maccallum}. These authors assume the Lewis form of the metric, where a cross term $dz\,d\phi$ would be absent. We did not find it possible
to bring (\ref{lambda}) to the Lewis form. The theorem by Papapetrou \cite{papapetrou} that any solution with two commuting Killing vectors (one timelike, one spacelike with periodic orbits) can be written in the Lewis form, only applies to solutions of the vacuum Einstein equations without cosmological term. Hence one cannot make a connection with spaces of the Lewis form as was possible for the Petrov solution.

\setcounter{equation}{0}
\section{Left-Invariant Forms and Non-Linear
  Realisations}
For a given homogeneous spacetime, one can consider the Lie algebra of its isometry group, i.e. the tangent space at the identity element, and construct left-invariant vector fields on the group by applying the push-forward of the left-translation to elements of the tangent space. The group structure of the manifold means that there is a global frame field of left-invariant vector fields (under the action of the group on itself).
\\In case of a matrix Lie group one conveniently uses the {\it Maurer-Cartan one-form} to obtain a basis of left-invariant covector fields (one-forms). The Maurer-Cartan form is defined by ($L_g$ denotes left-translation)
\bequ
\omega_g(v)=(L_{g^{-1}})_* v=(g^{-1}dg)(v)
\eequ
and by definition of a left-invariant vector field, applying $\omega$ to it returns the value of the vector field at the identity. If $\{e_a\}$ is a basis of the Lie algebra which induces a basis of left-invariant vector fields, and $\{\lambda^a\}$ is a dual basis of left-invariant one-forms, the Maurer-Cartan form can, by this reasoning, be written as
\bequ
\omega_g=g^{-1}dg=e_a\lambda^a.
\eequ
Direct computation of $g^{-1}dg$ in a matrix representation gives a basis of left-invariant one-forms.
\\
\\From a basis of left-invariant one-forms, one can construct general actions on a spacetime $X=G/K$, where $H\subset K$ is the stabilizer of a point, using non-linear realisations. A first order action has the general form
\bequ
S=\int (\alpha_i\lambda^i),
\eequ
where $\alpha_i$ are constants. The method was successfully applied to the construction of brane and superbrane actions in \cite{superbrane}, where one uses
\[G=E(n-1,1),\quad K=SO(n-1,1),\quad H=SO(p,1)\times SO(n-1-p)\]
for a $p$-brane, where $H$ consists of unbroken Lorentz rotations. The case $p=0$ will yield a point-particle action.
\\
\\As an explicit example, consider the group $SU(2)$ parametrised by Euler angles ($\varphi,\vartheta,\psi$)
\[g=e^{\psi D_z}e^{\vartheta D_x}e^{\varphi D_z}\]
where $D_z$ and $D_x$ generate rotations about the $z$- and $x$-axis, respectively\footnote{Since the Lie algebras of $SO(3)$ and $SU(2)$ are identical, we think of these as generating rotations in three real dimensions.}. We are looking  for non-linear realisations with $G=SU(2)$ and $H=K=id$. By calculating the Maurer-Cartan form one obtains the left-invariant forms
\bequ
\lambda^1=\cos\varphi\, d\vartheta+\sin\vartheta\sin\varphi\, d\psi,\;\lambda^2=\sin\varphi\, d\vartheta-\sin\vartheta\cos\varphi\, d\psi,\;\lambda^3=d\varphi+\cos\vartheta\, d\psi.
\eequ
A general action is then given by (here a dot denotes a derivative with respect to $\lambda$)
\bequ
S=\int d\lambda\;\left\{\alpha_1(\cos\varphi \dot{\vartheta}+\sin\vartheta\sin\varphi \dot{\psi})+\alpha_2(\sin\varphi \dot{\vartheta}-\sin\vartheta\cos\varphi \dot{\psi})+\alpha_3(\dot{\varphi}+\cos\vartheta \dot{\psi})\right\}.
\eequ
Since $\alpha_3\dot{\varphi}$ is a total derivative, it can be removed from the action. Then the field $\varphi$ is non-dynamical and can be removed using
\bequ
0=\frac{\partial\mathcal{L}}{\partial\varphi}=\sin\varphi(-\alpha_1\dot{\vartheta}+\alpha_2\sin\vartheta\dot{\psi})+\cos\varphi(\alpha_1\sin\vartheta\dot{\psi}+\alpha_2  \dot{\vartheta})\;\Rightarrow\;\tan\varphi=\frac{\alpha_1\sin\vartheta\dot{\psi}+\alpha_2  \dot{\vartheta}}{\alpha_1\dot{\vartheta}-\alpha_2\sin\vartheta\dot{\psi}}.
\eequ
The action becomes
\bequ
S=\int d\lambda\;\frac{\pm(\alpha_1\dot{\vartheta}-\alpha_2\sin\vartheta\dot{\psi})^2\pm(\alpha_2\dot{\vartheta}+\alpha_1\sin\vartheta\dot{\psi})^2}{\sqrt{(\alpha_1^2+\alpha_2^2)\dot{\vartheta}^2+(\alpha_1^2+\alpha_2^2)\sin^2\vartheta\dot{\psi}^2}}+\alpha_3\cos\vartheta \dot{\psi},
\eequ
and by choosing the positive signs
\bequ
S=\int d\lambda\;\sqrt{(\alpha_1^2+\alpha_2^2)}\sqrt{\dot{\vartheta}^2+\sin^2\vartheta\dot{\psi}^2}+\alpha_3\cos\vartheta \dot{\psi}.
\eequ
This action describes the motion of a point-particle on $S^2=SU(2)/U(1)$ with a magnetic moment, in the field of a magnetic monopole.\footnote{We owe this example to Joaquim Gomis.}

\subsection*{Application to the Petrov Spacetime}
Since the action of the elements of the group manifold on itself has already been described in the introductory section, we can write down a matrix representation of this group of motions, with a general element given by
\bequ
g=\left(\matrix{1&0&0&0&r \cr 0&e^{r}&0&0&z \cr
  0&0&e^{-\frac{r}{2}}\cos(\frac{\sqrt{3}}{2}r)&e^{-\frac{r}{2}}\sin(\frac{\sqrt{3}}{2}r)&\phi
  \cr
  0&0&-e^{-\frac{r}{2}}\sin(\frac{\sqrt{3}}{2}r)&e^{-\frac{r}{2}}\cos(\frac{\sqrt{3}}{2}r)&t
  \cr 0&0&0&0&1}\right).
\label{paramet}
\eequ
The group is generated by
\bequ
Z=\left(\matrix{0&0&0&0&0 \cr 0&0&0&0&1 \cr 0&0&0&0&0 \cr 0&0&0&0&0
  \cr 0&0&0&0&0}\right),\;\Phi=\left(\matrix{0&0&0&0&0 \cr 0&0&0&0&0 \cr 0&0&0&0&1 \cr 0&0&0&0&0
  \cr 0&0&0&0&0}\right),
\eequ
\[T=\left(\matrix{0&0&0&0&0 \cr 0&0&0&0&0 \cr 0&0&0&0&0 \cr 0&0&0&0&1
  \cr 0&0&0&0&0}\right),R=\left(\matrix{0&0&0&0&1 \cr 0&1&0&0&0 \cr 0&0&-\frac{1}{2}&\frac{\sqrt{3}}{2}&0 \cr 0&0&-\frac{\sqrt{3}}{2}&-\frac{1}{2}&0
  \cr 0&0&0&0&0}\right),\]
so that $g=e^{zZ+\phi\Phi+tT}e^{rR}$ and $(r,z,\phi,t)$ are coordinates on the group. The generators satisfy the algebra
\bequ
[R,T]=-\frac{1}{2}T+\frac{\sqrt{3}}{2}\Phi,\;
  [R,\Phi]=-\frac{1}{2}\Phi-\frac{\sqrt{3}}{2}T,\; [R,Z]=Z.
\label{genalg}
\eequ
This differs from the Killing algebra (\ref{algebra}) by the usual overall minus sign coming from the fact that right-invariant vector fields generate left actions and vice versa. 
\\The matrix representation (\ref{paramet}) gives the group multiplication law
\bequ
(r,z,\phi,t)\cdot(r',z',\phi',t')=\left(r+r',z+e^r z',\phi+e^{-\frac{r}{2}}(t's+\phi'c),t+e^{-\frac{r}{2}}(t'c-\phi's)\right),
\label{multlaw}
\eequ
where $s\equiv\sin\frac{\sqrt{3}}{2}r,\;c\equiv\cos\frac{\sqrt{3}}{2}r$. The Maurer-Cartan form is
\bequ
g^{-1}dg=e^{-rR}(Zdz+\Phi d\phi+Tdt)e^{rR}+drR=R\,dr+
\label{maurer}
\eequ
\[e^{-r}Zdz+e^{\frac{r}{2}}\left(\cos\left(\frac{\sqrt{3}}{2}r\right)\Phi+\sin\left(\frac{\sqrt{3}}{2}r\right)T\right)d\phi+e^{\frac{r}{2}}\left(-\sin\left(\frac{\sqrt{3}}{2}r\right)\Phi+\cos\left(\frac{\sqrt{3}}{2}r\right)T\right)dt\]
which gives the desired basis of left-invariant one-forms:
\bequ
\lambda^1=dr,\;\lambda^2=e^{-r}\,dz,
\label{1forms}
\eequ
\[\lambda^3=e^{\frac{r}{2}}\left(\cos\left(\frac{\sqrt{3}}{2}r\right)\,d\phi-\sin\left(\frac{\sqrt{3}}{2}r\right)\,dt\right),\;\lambda^4=e^{\frac{r}{2}}\left(\sin\left(\frac{\sqrt{3}}{2}r\right)\,d\phi+\cos\left(\frac{\sqrt{3}}{2}r\right)\,dt\right).\]
We obtain a left-invariant metric on the group
\[ds^2=\eta_{\mu\nu}\lambda^{\mu}\otimes\lambda^{\nu}=dr^2+e^{-2r}dz^2+e^{r}\left(\cos(\sqrt{3}r)(d\phi^2-dt^2)-2\sin(\sqrt{3}r)d\phi\,dt\right)\]
with $\eta=$ diag$(1,1,1,-1)$, which is the same as (\ref{petrov}) and shows that our chosen coordinates agree with the initial Petrov coordinates. We see how to recover a metric on a group manifold; note that one could obtain this metric by just starting from the algebra (\ref{algebra}).
\\
\\A general (first-order) action would have the form
\bequ
S=\int d\lambda\;\left[\alpha\dot{r}+\beta e^{-r}\dot{z}+\gamma e^{\frac{r}{2}}\left(\cos\left(\frac{\sqrt{3}}{2}r\right)\,\dot{\phi}-\sin\left(\frac{\sqrt{3}}{2}r\right)\,\dot{t}\right)\right.
\label{action}
\eequ
\[\left.+\delta e^{\frac{r}{2}}\left(\sin\left(\frac{\sqrt{3}}{2}r\right)\,\dot{\phi}+\cos\left(\frac{\sqrt{3}}{2}r\right)\,\dot{t}\right)\right]\]
Since $\alpha\dot{r}$ is a total derivative, an action without this term is equivalent to the given one. Then the Lagrangian will not contain $\dot{r}$ and hence $r$ is a non-dynamical field corresponding to a Lagrange multiplier. Removing it from the action will lead to an action on a torus $T^3$. However,
\bequ
0=\frac{\partial\mathcal{L}}{\partial r}=-\beta e^{-r}\dot{z}+e^{\frac{r}{2}}\cos\left(\frac{\sqrt{3}}{2}r\right)\left(\frac{1}{2}(\gamma+\sqrt{3}\delta)\dot{\phi}+\frac{1}{2}(\delta-\sqrt{3}\gamma)\dot{t}\right)
\label{unsolvable}
\eequ
\[+e^{\frac{r}{2}}\sin\left(\frac{\sqrt{3}}{2}r\right)\left(\frac{1}{2}(\delta-\sqrt{3}\gamma)\dot{\phi}-\frac{1}{2}(\gamma+\sqrt{3}\delta)\dot{t}\right)\]
can not be solved algebraically for $r$ and hence we are not able to give an explicit form of the action without $r$. But although we have been unable to find an explicit Lagrangian depending only on $\dot{z},\dot{\phi},\dot{t}$ by eliminating $r$, we can see that the result must be homogeneous of degree one (but highly non-linear) in $\dot{z},\dot{\phi}$ and $\dot{t}$. This is because (\ref{unsolvable}) is unchanged under the rescaling $(\dot{z},\dot{\phi},\dot{t})\rightarrow (\lambda\dot{z},\lambda\dot{\phi},\lambda\dot{t})$ with $\lambda\neq 0$. As a consequence $r$ is a homogeneous function of degree zero in $(\dot{z},\dot{\phi},\dot{t})$. It follows that if we now substitute $r=r(\dot{z},\dot{\phi},\dot{t})$ into (\ref{action}) the result will be homogeneous of degree one in $(\dot{z},\dot{\phi},\dot{t})$. The resulting Lagrangian is of course highly non-linear but it is of Finsler type.
\\It is striking that in this case, as in that discussed in \cite{finsler}, that the method of non-linear realisations leads to a Lagrangian of Finsler type.
\\
\\The difficulties encountered here suggest that the method of non-linear realisations in general does not always give an explicit implementation of arbitrary given isometries.

\setcounter{equation}{0}
\section{Deformations of Lie Algebras}
A mathematical operation of interest in the context of Lie algebras is the {\it deformation} of a given Lie algebra \cite{levy-n}. One can describe the action of the Lie bracket on the Lie algebra (with basis $\{e_a\}$) by {\it structure constants} ${{C_a}^c}_b$, defined by
\[[e_a,e_b]={{C_a}^c}_b e_c.\]
Then consider the algebraic manifold formed by the set of the possible collections of structure constants of the Lie algebra  and view a deformation of the Lie algebra as a curve in this manifold:
\[{{\hat{C}_a}\,^c}_b (t)={{C_a}^c}_b+t{{A_a}^c}_b+t^2{{B_a}^c}_b+\ldots\]
The manifold is defined by the Jacobi equation, which in terms of the structure constants means that for each $t$
\bequ
{{\hat{C}_d}\,^e}_{[a}(t){{\hat{C}_b}\,^d}_{c]}(t)=0,
\label{jacobi}
\eequ
which at linear order leads to the requirement
\bequ
{{C_d}^e}_{[a}{{A_b}^d}_{c]}+{{A_d}^e}_{[a}{{C_b}^d}_{c]}=0.
\label{jaclin}
\eequ
A linear deformation only gives rises to a deformation if it is {\it integrable}, if the requirement (\ref{jacobi}) can be satisfied at each order in $t$.
\\A deformation corresponding to a change of basis, i.e. a linear map $S$ acting on the Lie algebra, such that $[a,b]^*=S[S^{-1}a,S^{-1}b]$, for instance, will be regarded as trivial. In this case the structure constants will change according to
\bequ
{{\hat{C}_a}\,^b}_c (t)={S^b}_e{{C_d}^e}_f{(S^{-1})^d}_a{(S^{-1})^f}_c.
\eequ
Expanding ${S^a}_b(t)=\delta^a_b+t{M^a}_b+\ldots$, this means that to first order a trivial deformation can be written as
\bequ
{{A_a}^b}_c={M^b}_e{{C_a}^e}_c-{{C_e}^b}_c{M^e}_a-{{C_a}^b}_e{M^e}_c.
\label{nontriv}
\eequ
The requirements (\ref{jaclin}) and (\ref{nontriv}) can be rephrased in the language of differential forms \cite{finsler}; a basis $\{\lambda^a\}$ of left-invariant one-forms for the original algebra satisfies $d\lambda^a=-\frac{1}{2}{{C_b}^a}_c\lambda^b\wedge\lambda^c$; define $M^a={M^a}_b\lambda^b$ and $A^a=\frac{1}{2}{{A_b}^a}_c\lambda^b\wedge\lambda^c$ to be vector-valued one- and two-forms and ${C^a}_b={{C_c}^a}_b\lambda^c$ to be a matrix-valued one-form. Then (\ref{jaclin}) and (\ref{nontriv})  can be rewritten as
\[DA=0,\;A\neq DM,\quad\mbox{where }D=d+C\wedge.\]
Therefore one can determine all general deformations of a given algebra by using cohomology theory \cite{farrill}. We will restrict ourselves to linear deformations.

\subsection*{Deformations of the Petrov Killing Algebra}
We examine possible
deformations of the four-dimensional Lie algebra. Equations
(\ref{jaclin}) give the following conditions on linear deformations:
\bequ
0={{A_z}^r}_{\phi}+\sqrt{3}{{A_t}^r}_z={{A_z}^r}_t-\sqrt{3}{{A_{\phi}}^r}_z={{A_{\phi}}^r}
_t;
\label{constr}
\eequ
\[0=2{{A_{\phi}}^r}_r+{{A_{\phi}}^z}_z+\sqrt{3}{{A_t}^z}_z=-\sqrt{3}{{A_{\phi}}^z}_z+2{{A_t}^r}_r+{{A_t}^z}_z={{A_{\phi}}^z}_t={{A_{\phi}}^r}_t;\]
\[0={{A_r}^r}_z+2{{A_{\phi}}^{\phi}}_z+\sqrt{3}{{A_z}^t}_{\phi}-\sqrt{3}{{A_t}^{\phi}}_z=\sqrt{3}{{A_r}^r}_z+2{{A_z}^{\phi}}_t-\sqrt{3}{{A_z}^t}_t-\sqrt{3}{{A_{\phi}}^{\phi}}_z\]
\[=\sqrt{3}{{A_r}^r}_{\phi}+{{A_t}^{\phi}}_{\phi}-\sqrt{3}{{A_{\phi}}^t}_t-{{A_t}^r}_r=\sqrt{3}{{A_z}^r}_{\phi}-{{A_t}^r}_z;\]
\[0=\sqrt{3}{{A_r}^r}_z-\sqrt{3}{{A_z}^{\phi}}_{\phi}-2{{A_z}^t}_{\phi}-\sqrt{3}{{A_t}^t}_z={{A_r}^r}_z-\sqrt{3}{{A_z}^{\phi}}_t+2{{A_t}^t}_z+\sqrt{3}{{A_{\phi}}^t}_z\]
\[={{A_r}^r}_{\phi}-\sqrt{3}{{A_{\phi}}^{\phi}}_t+\sqrt{3}{{A_t}^r}_r-{{A_t}^t}_{\phi}={{A_z}^r}_{\phi}+\sqrt{3}{{A_t}^r}_z.\]
These constraints for ${{A_a}^c}_b$ reduce the number of free parameters from 24 to twelve. We list the most general deformation parameters satisfying (\ref{constr}) in a table:
\begin{center}
\begin{tabular}{c|c|c|c|c}
& $c=r$&$c=z$&$c=\phi$&$c=t$
\\\hline
${{A_r}^c}_z$&$2C$&$x_1$&$x_2$&$x_3$
\\\hline
${{A_r}^c}_{\phi}$&$-\sqrt{3}A-B$&$x_4$&$x_5$&$x_6$
\\\hline
${{A_r}^c}_t$&$-A+\sqrt{3}B$&$x_7$&$x_8$&$x_9$
\\\hline
${{A_z}^c}_{\phi}$&0&$2B$&$C$&$\sqrt{3}C$
\\\hline
${{A_z}^c}_t$&0&$2A$&$-\sqrt{3}C$&$C$
\\\hline
${{A_{\phi}}^c}_t$&0&0&$-\sqrt{3}B-A$&$B-\sqrt{3}A$
\end{tabular}
\end{center}
The parameters $x_1,x_2,\ldots,x_9,A,B$ and $C$ can be arbitrary real constants. We need to investigate which of these correspond to trivial deformations. The conditions (\ref{nontriv}) mean that trivial deformations can be written as
\bequ
{{A_r}^r}_z=-{M^r}_z,\;{{A_r}^z}_z=-{M^r}_r,\;{{A_r}^{\phi}}_z=-\frac{3}{2}{M^{\phi}}_z+\frac{\sqrt{3}}{2}{M^t}_z,\;{{A_r}^t}_z=-\frac{\sqrt{3}}{2}{M^{\phi}}_z-\frac{3}{2}{M^t}_z.
\eequ
The parameters $C,x_1,x_2$ and $x_3$ correspond to trivial deformations and can be set to zero by a change of basis. Furthermore,
\bequ
{{A_r}^r}_{\phi}=\frac{1}{2}{M^r}_{\phi}+\frac{\sqrt{3}}{2}{M^r}_t,\;{{A_r}^r}_t=-\frac{\sqrt{3}}{2}{M^r}_{\phi}+\frac{1}{2}{M^r}_t,\;{{A_z}^z}_{\phi}=-{M^r}_{\phi},\;{{A_z}^z}_t=-{M^r}_t
\eequ
etc., so that we can set $A=B=0$,
\bequ
{{A_r}^z}_{\phi}=\frac{3}{2}{M^z}_{\phi}+\frac{\sqrt{3}}{2}{M^z}_t,\;{{A_r}^{\phi}}_{\phi}=-\frac{1}{2}{M^r}_r+\frac{\sqrt{3}}{2}{M^{\phi}}_t+\frac{\sqrt{3}}{2}{M^t}_{\phi},\;{{A_r}^t}_{\phi}=-\frac{\sqrt{3}}{2}{M^r}_r-\frac{\sqrt{3}}{2}{M^{\phi}}_{\phi}+\frac{\sqrt{3}}{2}{M^t}_t,
\eequ
so that we can set $x_4=x_5=x_6=0$,
\bequ
{{A_r}^z}_t=\frac{3}{2}{M^z}_t-\frac{\sqrt{3}}{2}{M^z}_{\phi},\;{{A_r}^{\phi}}_t=\frac{\sqrt{3}}{2}{M^r}_r-\frac{\sqrt{3}}{2}{M^{\phi}}_{\phi}+\frac{\sqrt{3}}{2}{M^t}_t,\;{{A_r}^t}_t=-\frac{1}{2}{M^r}_r-\frac{\sqrt{3}}{2}{M^{\phi}}_t-\frac{\sqrt{3}}{2}{M^t}_{\phi},
\eequ
so that we can set $x_7=0$, but must treat $x_8$ and $x_9$ as nontrivial perturbations. After a relabelling of these parameters, the modified Lie algebra is now
\bequ
[R,T]=-aT-b\Phi,\;  [R,\Phi]=-\frac{1}{2}\Phi-\frac{\sqrt{3}}{2}T,\; [R,Z]=Z.
\eequ
These relations satisfy the full Jacobi identities and so the linear deformation indeed defines a deformation of the Lie algebra. We may modify the matrix representation by setting
\bequ
R=\left(\matrix{0&0&0&0&1 \cr 0&1&0&0&0 \cr 0&0&-\frac{1}{2}&-b&0 \cr 0&0&-\frac{\sqrt{3}}{2}&-a&0
  \cr 0&0&0&0&0}\right).
\eequ
\\In the case where $a\neq\frac{1}{2}$, one can always find a linear transformation of the basis vectors $\Phi$
and $T$ such that the algebra takes the more symmetric form
\bequ
[R,T]=a'T+b'\Phi,\;  [R,\Phi]=a'\Phi\pm b'T,\; [R,Z]=Z,
\eequ
with $\pm$ depending on the value of $b$ in the original deformation. This means that there
are three distinct cases: 
\\{\bf First case: Positive sign.} Then a matrix representation of
$R$ is
\bequ
R=\left(\matrix{0&0&0&0&1 \cr 0&1&0&0&0 \cr 0&0&a'&b'&0 \cr 0&0&b'&a'&0
  \cr 0&0&0&0&0}\right)
\eequ
and a general group element looks like
\bequ
g=e^{zZ+\phi\Phi+tT}e^{rR}=\left(\matrix{1&0&0&0&r \cr 0&e^{r}&0&0&z \cr
  0&0&e^{a' r}\cosh(b' r)&e^{a' r}\sinh(b' r)&\phi
  \cr
  0&0&e^{a' r}\sinh(b' r)&e^{a' r}\cosh(b' r)&t
  \cr 0&0&0&0&1}\right).
\eequ
The Maurer-Cartan form is
\bequ
g^{-1}dg=Z\,e^{-r}dz+\Phi (e^{-ar}(\cosh(b'r)d\phi-\sinh(b'r)dt)+T (e^{-ar}(\cosh(b'r)dt-\sinh(b'r)d\phi)+R\, dr
\eequ
and a left-invariant metric will be given by
\[ds^2=\eta_{\mu\nu}\lambda^{\mu}\otimes\lambda^{\nu}=dr^2+e^{-2r}dz^2+e^{-2a'r}(d\phi^2-dt^2).\]
The special case $a'=1,\;b'=3$ gives the algebra of the Kaigorodov-Ozsv\'ath solution (\ref{lambda}), as can be seen by setting $\Phi'=\Phi-T$ and $T'=\Phi+T$, which amounts to
\bequ
[R,\Phi]=\Phi+3T,\;[R,T]=3\Phi+T\quad\Rightarrow\;[R,\Phi']=-2\Phi',\;[R,T']=4T',
\eequ
which is just the sign-reversed version of (\ref{algnew}). One can recover the metric (\ref{lambda}) by choosing the symmetric matrix
\bequ
h_{\mu\nu}=\left(\matrix{-\frac{3}{\Lambda}&0&0&0\cr 0&1&-\sqrt{2}&\sqrt{2} \cr 0&-\sqrt{2}&-1&-1\cr 0&\sqrt{2}&-1&3}\right)
\label{newmatrix}
\eequ
and computing the left-invariant metric
\bequ
h_{\mu\nu}\lambda^{\mu}\otimes\lambda^{\nu}=-\frac{3}{\Lambda}dr^2+e^{-2r}dz^2-2e^{-2r}(d\phi^2-dt^2)+e^{4r}(d\phi-dt)^2-2\sqrt{2}e^r dz(d\phi-dt),
\eequ
which after the coordinate transformations $\phi-t=\phi'$ and $-\phi-t=t'$ reduces to (\ref{lambda}). Since the matrix (\ref{newmatrix}) has one negative and three positive eigenvalues, there exists a (vierbein) basis of left-invariant one-forms $\sigma^{\mu}$ such that $\eta_{\mu\nu}\sigma^{\mu}\otimes\sigma^{\nu}$ gives the metric (\ref{lambda}).
\\
\\A general first-order action on the Kaigorodov-Ozsv\'ath spacetime would have the form
\bequ
S=\int d\lambda\;\left[\alpha\dot{r}+e^{-r}\left(\beta \dot{z}+\gamma \left(\cosh(3r)\,\dot{\phi}-\sinh(3r)\,\dot{t}\right)+\delta \left(-\sinh(3r)\,\dot{\phi}+\cosh(3r)\,\dot{t}\right)\right)\right].
\label{akschn}
\eequ
As before, $r$ is a non-dynamical field corresponding to a Lagrange multiplier. One could remove it from the action by solving
\bequ
0=\frac{\partial\mathcal{L}}{\partial r}=-e^{-r}\left(\beta \dot{z}+\cosh(3r)((\gamma+3\delta)\dot\phi+(\delta+3\gamma)\dot t)-\sinh(3r)((\gamma+3\delta)\dot t+(\delta+3\gamma)\dot\phi)\right)
\eequ
for $r$, which can in this case be done explicitly; we obtain
\bequ
r=\frac{1}{3}\log\left(\frac{\beta\dot{z}+\sqrt{\beta^2\dot{z}^2+8(\dot{\phi}^2-\dot{t}^2)(\gamma^2-\delta^2)}}{2(\dot\phi-\dot t)(\gamma-\delta)}\right)
\eequ
and hence the action (\ref{akschn}) becomes
\bequ
S=\int d\lambda\;\left(\frac{2(\dot\phi-\dot t)(\gamma-\delta)}{\beta\dot{z}+\sqrt{\beta^2\dot{z}^2+8(\dot{\phi}^2-\dot{t}^2)(\gamma^2-\delta^2)}}\right)^{1/3}\left\{\frac{5}{4}\beta\dot{z}+\frac{1}{4}\sqrt{\beta^2\dot{z}^2+8(\dot{\phi}^2-\dot{t}^2)(\gamma^2-\delta^2)}\right.
\eequ
\[\left.+\frac{(\dot\phi-\dot t)(\gamma-\delta)}{\beta\dot{z}+\sqrt{\beta^2\dot{z}^2+8(\dot{\phi}^2-\dot{t}^2)(\gamma^2-\delta^2)}}\right\}.\]
This action is still homogeneous of degree one since the expression for $r$ is homogeneous of degree zero; again we obtain a Lagrangian of Finsler type, just as in the case of the Petrov spacetime. Since this new action, just as the original one, does not depend on $z,\phi$ or $t$, the associated conjugate momenta are conserved quantities and the theory is integrable.
\\
\\{\bf Second case: Negative sign.}
\bequ
R=\left(\matrix{0&0&0&0&1 \cr 0&1&0&0&0 \cr 0&0&a'&b'&0 \cr 0&0&-b'&a'&0
  \cr 0&0&0&0&0}\right),\;
g=e^{zZ+\phi \Phi+tT}e^{rR}=\left(\matrix{1&0&0&0&r \cr 0&e^{r}&0&0&z \cr
  0&0&e^{a' r}\cos(b' r)&e^{a' r}\sin(b' r)&\phi
  \cr
  0&0&-e^{a' r}\sin(b' r)&e^{a' r}\cos(b' r)&t
  \cr 0&0&0&0&1}\right).
\eequ
The Maurer-Cartan form is
\bequ
g^{-1}dg=Z\,e^{-r}dz+\Phi (e^{-ar}(\cos(b'r)d\phi-\sin(b'r)dt)+T (e^{-ar}(\cos(b'r)dt+\sin(b'r)d\phi)+R\, dr
\eequ
and a left-invariant metric will be given by
\bequ
ds^2=\eta_{\mu\nu}\lambda^{\mu}\otimes\lambda^{\nu}=dr^2+e^{-2r}dz^2+e^{-2a'r}(\cos(2b'r)(d\phi^2-dt^2)-2\sin(2b'r)d\phi\,dt).
\label{linv}
\eequ
The original Petrov algebra is of course the special case $a'=-\half,\;b'=\frac{\sqrt{3}}{2}$.
\\For the metric (\ref{linv}) the Ricci tensor has non-vanishing components
\[R_{rr}=-1-2a'^2+2b'^2,\;R_{zz}=-(1+2a')e^{-2r},\;R_{\phi\phi}=-(1+2a')e^{-2a'r}(a'\cos(2b'r)+b'\sin(2b'r)),\]
\[R_{\phi t}=(1+2a')e^{-2a'r}(a'\sin(2b'r)-b'\cos(2b'r)),\;R_{tt}=(1+2a')e^{-2a'r}(a'\cos(2b'r)+b'\sin(2b'r)).\]
The manifold is an Einstein manifold only if $a'=-\half$ and $b'=\pm\frac{\sqrt{3}}{2}$ ($\Lambda=0$) or $a'=1$ and $b'=0$ ($\Lambda=-3$, anti-de Sitter space). In the general case the energy-momentum tensor defined by $T_{ab}=\frac{1}{8\pi G}G_{ab}$ does not satisfy the weak energy condition; without loss of generality assume $\sin(\sqrt{3}r)=0$ and $\cos(\sqrt{3}r)=1$ and choose a timelike vector $t^a=(0,0,t_3,t_4)$ ($t_4^2\ge t_3^2$), then
\bequ
G_{ab}t^a t^b=(1+a'+a'^2-b'^2)(t_3^2-t_4^2)-2(1+2a')b't_3t_4
\eequ
can be made arbitrarily negative by letting $t_3,t_4\rightarrow\pm\infty$ while keeping $t_4^2-t_3^2$ small and positive, unless $a'=-\half$ and $b'^2\ge \frac{3}{4}$. In the case $a'=-\half$ and $b'^2\ge \frac{3}{4}$, the Einstein tensor can be written as
\bequ
G_{ab}=-\lambda g_{ab}+2\lambda u_a u_b,\quad\lambda\equiv b'^2-\frac{3}{4}\ge 0,\quad u_a=(1,0,0,0),\;u_a u^a=1.
\label{fluid}
\eequ
This would resemble the energy-momentum tensor of a perfect fluid for a timelike vector $u_a$, but here $u_a$ is spacelike. Hence we can not give an obvious physical interpretation to this spacetime. 
\\The Einstein tensor defined by (\ref{fluid}) satisfies the dominant energy condition as $2t^r v^r-t^a v_a\ge 0$ for any timelike and future-directed $t,v$. These statements are independent of the choice of the arrow of time, i.e. hold for both $t_4<0$ or $t_4>0$.
\\
\\{\bf Third case: $a=\frac{1}{2}$}. Let us introduce a new parameter $c$, so that
\bequ
[R,T]=-\frac{1}{2}T-\frac{2}{\sqrt{3}}c^2 \Phi,\;
  [R,\Phi]=-\frac{1}{2}\Phi-\frac{\sqrt{3}}{2}T,\; [R,Z]=Z
\eequ
for positive $b$ which gives
\[R=\left(\matrix{0&0&0&0&1 \cr 0&1&0&0&0 \cr 0&0&-\frac{1}{2}&-\frac{2c^2}{\sqrt{3}}&0 \cr 0&0&-\frac{\sqrt{3}}{2}&-\frac{1}{2}&0
  \cr 0&0&0&0&0}\right)\]
and a general group element looks like
\bequ
g=e^{zZ+\phi\Phi+tT}e^{rR}=\left(\matrix{1&0&0&0&r \cr 0&e^{r}&0&0&z \cr
  0&0&e^{-\frac{r}{2}}\cosh(c r)&-\frac{2c}{\sqrt{3}}e^{-\frac{ r}{2}}\sinh(c r)&\phi
  \cr
  0&0&-\frac{\sqrt{3}}{2c}e^{-\frac{ r}{2}}\sinh(c r)&e^{-\frac{ r}{2}}\cosh(c r)&t
  \cr 0&0&0&0&1}\right).
\eequ
The Maurer-Cartan form is
\bequ
g^{-1}dg=R\,dr+Z\,e^{-r}\,dz+\Phi\left(e^{\frac{r}{2}}\cosh(cr)d\phi+\frac{2c}{\sqrt{3}}e^{\frac{r}{2}}\sinh(cr)dt\right)+T\left(e^{\frac{r}{2}}\cosh(cr)dt+\frac{\sqrt{3}}{2c}e^{\frac{r}{2}}\sinh(cr)d\phi\right).
\eequ
In the limit $c\rightarrow 0$ or $b\rightarrow 0$ this becomes
\bequ
g^{-1}dg=R\,dr+Z\,e^{-r}\,dz+\Phi\,e^{\frac{r}{2}}d\phi+T\left(e^{\frac{r}{2}}dt+\frac{\sqrt{3}}{2}e^{\frac{r}{2}}r\,d\phi\right).
\eequ
The only remaining case, namely negative $b$, is
\bequ
[R,T]=-\frac{1}{2}T+\frac{2}{\sqrt{3}}c^2 \Phi,\;
  [R,\Phi]=-\frac{1}{2}\Phi-\frac{\sqrt{3}}{2}T,\; [R,Z]=-Z,
\eequ
which turns the hyperbolic into trigonometric functions. 
\\If $c\neq 0$ one can rescale the coordinate $t$ (for instance) and recover the same left-invariant forms as before, hence this does not give anything new. In the case $c=0$ a left-invariant metric is given by
\bequ
ds^2=dr^2+e^{-2r}dz^2+e^r\left(\left(1-\frac{3}{4}r^2\right)d\phi^2-\sqrt{3}r\,d\phi\,dt-dt^2\right).
\eequ
The Einstein tensor for this metric can be written as
\bequ
G_{ab}=\frac{3}{16} g_{ab}+\frac{3}{8}\left(\mbox{diag}\left(-2,e^{-2r},2e^r,0\right)\right)_{ab}.
\eequ
This does not satisfy the weak energy condition as $G_{ab}t^a t^b=-\frac{3}{16}(3+e^r)<0$ for the timelike vector $t^a=(1,0,0,1)$.

\setcounter{equation}{0}
\section{Spinning Cylinders}
In the case of an infinite rigidly rotating dust cylinder one could imagine trying to speed up this cylinder by shooting in particles with some angular momentum which enter the interior region on causal curves and increase the angular velocity $a$, so as to reach and surpass the critical value $aR=\half$ above which CTCs appear. We will show that this is not possible.
\\The interior part of the general van Stockum solution
\bequ
ds^2=H(\rho)(d\rho^2+dz^2)+L(\rho)d\phi^2+2M(\rho)d\phi\, dt-F(\rho)dt^2,
\eequ
describing the region $\rho<R$ is given by \cite{tipler74}
\bequ
H=\exp(-a^2 \rho^2),\;L=\rho^2(1-a^2 \rho^2),\;M=a\rho^2,\;F=1.
\eequ
As there are closed timelike curves for $\rho>\frac{1}{a}$ we require $aR\le 1$. The exterior solution is for $a<\frac{1}{2}$, from now on setting $R=1$ for simplicity which is no loss of generality,
\bequ
H=e^{-a^2}\rho^{-2a^2},\;L=\frac{\rho\sinh(3\epsilon+\theta)}{2\sinh 2\epsilon \cosh\epsilon},\;M=\frac{\rho\sinh(\epsilon+\theta)}{\sinh 2\epsilon},\;F=\frac{\rho\sinh(\epsilon-\theta)}{\sinh\epsilon}
\eequ
where $\theta(\rho)=\sqrt{1-4a^2}\log\rho$ and $\epsilon=$Artanh$\sqrt{1-4a^2}$. Note that always $-FL-M^2=-\rho^2$ and so the metric has the right signature for all $\rho$ (this of course is also true as $a\rightarrow\half$). The point-particle Lagrangian is (a dot denotes differentiation with respect to an affine parameter $\lambda$)
\bequ
\mathcal{L}=g_{ij}\dot{x}^i\dot{x}^j=H(\dot{\rho}^2+\dot{z}^2)+L\dot{\phi}^2+2M\dot{\phi}\dot{t}-F\dot{t}^2,
\eequ
and since the Lagrangian does not depend on $z,\phi$ and $t$ there are three conserved quantities associated with geodesics:
\[P\equiv H\dot{z},\;J\equiv L\dot{\phi}+M\dot{t},\;E\equiv F\dot{t}-M\dot{\phi}.\]
The Lagrangian for timelike or null geodesics becomes
\bequ
\mathcal{L}=H\dot{\rho}^2+\frac{P^2}{H}+\frac{1}{\rho^2}\left(FJ^2-2MEJ-LE^2\right)=-m^2\le 0
\label{lagrarsch}
\eequ
and we obtain the radial equation
\bequ
\left(\frac{d\rho}{d\lambda}\right)^2=\frac{L}{H\rho^2}\left(-\frac{\rho^2
  m^2}{L}-\frac{\rho^2P^2}{HL}+E^2+\frac{2M}{L}EJ-\frac{F}{L}J^2\right)=\frac{L}{H\rho^2}(E-V_{eff}^+(\rho))(E-V_{eff}^-(\rho)),
\eequ
where we have introduced an effective potential
\bequ
V_{eff}^{\pm}(\rho)=\frac{M(\rho)}{L(\rho)}J\pm \rho\sqrt{\frac{1}{L(\rho)}\left(m^2+\frac{P^2}{H(\rho)}+\frac{J^2}{L(\rho)}\right)}.
\eequ
This is well-defined for all $\rho$ as $H,\;L,\;M$ all remain positive for all $\rho$. A particle falling in on a geodesic can enter the cylinder if 
\bequ
E>V_{eff}^+(1)=\frac{a}{1-a^2}J+ \sqrt{\frac{1}{1-a^2}\left(m^2+P^2 e^{a^2}+\frac{J^2}{1-a^2}\right)}\ge\frac{a+1}{1-a^2}J=\frac{1}{1-a}J.
\eequ
Any particle entering the cylinder on a geodesic must have $\frac{J}{E}<1-a.$ In the limit $a\rightarrow 0$ the conserved quantities $J$ and $E$ clearly describe angular momentum and energy per mass. We can identify the ratio $\frac{J}{E}$ with the angular velocity of an infalling particle at $R=1$.
\\If we are considering accelerated observers, equation (\ref{lagrarsch}) still holds, but $P,\;E$ and $J$ will no longer be conserved quantities. However, only the local values of these quantities at $R=1$ will decide about whether or not a particle will be able to enter the interior region of the cylinder.
\\
\\This means that the above considerations also hold for accelerated observers and as any particle entering the cylinder must have $\frac{J}{E}<\half$ for $a=\half$, one cannot speed up the cylinder beyond $a=\half$ using particles on timelike or null curves.

\setcounter{equation}{0}
\section{Geodesics}

\subsection{Geodesic Completeness}
We ask whether the Petrov spacetime, with the radial coordinate $r$ extended to take arbitrary values, is
geodesically complete, i.e. whether all timelike and null geodesics can be extended to
infinite values of the affine parameter. First we give an example that this need not be possible on a group manifold: 
Remove the null hyperplane $z=t$ from Minkowski space and consider the half-space $z>t$, denoted by $M^-$. It is clearly geodesically incomplete. Null translations and boosts
\bequ
(z,t)\rightarrow  (z+c,t+c);\quad (t+z,t-z)\rightarrow \left(\lambda(t+z),\frac{1}{\lambda}(t-z)\right)
\eequ
act on $M^-$, and together with translations $(x,y)\rightarrow(x+a,y+b)$ they form a four-dimensional group which acts simply-transitively on $M^-$. For instance, the point $(x,y,z,t)=(0,0,1,0)$ is, by a null translation and a successive boost, taken to
\bequ
(0,0,1,0)\rightarrow (0,0,1+c,c)\rightarrow \left(0,0,\lambda c+\half-\frac{1}{2\lambda},\lambda c+\half+\frac{1}{2\lambda}\right).
\eequ
There is a one-one correspondence between points in $M^-$ and group parameters $(a,b,\lambda,c)$, where $\lambda>0$. The space $M^-$ can be identified with the group $G\times \bR^2$, where $G$ is the unique two-dimensional non-Abelian Lie group.
\\As a second example, introduced in a slightly different context in \cite{polyakov}, consider the dilatation group generated by translations and dilatations
\bequ
x^a\rightarrow x^a+c^a,\quad x^a\rightarrow \rho x^a,
\eequ
where we denote the generators by $P_a$ and $D$ respectively. Parametrising the group elements by $g=e^{x^a P_a}e^{\lambda D}$, the Maurer-Cartan form is
\bequ
g^{-1}dg=e^{-\lambda}dx^a\,P_a+d\lambda\,D
\eequ
and hence a left invariant metric is
\bequ
ds^2=d\lambda^2+e^{-2\lambda}\eta_{ab}dx^a dx^b=\frac{1}{\rho^2}(d\rho^2+\eta_{ab}dx^a dx^b),
\eequ
where $\rho=e^{\lambda}$. This is the metric of anti-de Sitter space in five dimensions in Poincar\'e coordinates, which is geodesically incomplete as these coordinates cover only a patch of the full spacetime. Hence geodesic completeness is a non-trivial property of a group manifold.
\\
\\To show geodesic completeness of the Petrov solution, we need to show that no geodesic reaches infinity for finite values of the affine parameter. Consider the Lagrangian
\bequ
\mathcal{L}=g_{ab}\dot{x}^a\dot{x}^b,
\eequ
which for the metric (\ref{petrov}) is
\bequ
\mathcal{L}=\dot{r}^2+e^{-2r}\dot{z}^2+e^r\left(\cos\sqrt{3}r(\dot{\phi}^2-\dot{t}^2)-2\dot{\phi}\dot{t}\sin\sqrt{3}r\right).
\eequ
Evidently, from the Euler-Lagrange equations, there are three conserved
quantities because the Lagrangian does not depend on $z,\phi$ or $t$
explicitly, the conjugate momenta
\bequ
P\equiv e^{-2r}\dot{z},\;j\equiv
e^r(\dot{\phi}\cos\sqrt{3}r-\dot{t}\sin\sqrt{3}r),\;E\equiv e^r(\dot{\phi}\sin\sqrt{3}r+\dot{t}\cos\sqrt{3}r).
\eequ
The Lagrangian now takes the form
\bequ
\mathcal{L}=\dot{r}^2+e^{2r}P^2+e^{-r}\left(\cos\sqrt{3}r(j^2-E^2)+2Ej\sin\sqrt{3}r\right),
\eequ
and since the Lagrangian is a conserved quantity in geodesic motion the equation
\bequ
\left(\frac{dr}{d\lambda}\right)^2=e^{-r}(E^2\cos\sqrt{3}r-2Ej\sin\sqrt{3}r-j^2\cos\sqrt{3}r)-e^{2r}P^2-m^2
\label{radial}
\eequ
is satisfied by any geodesic, where $m^2$ is positive for timelike,
zero for null and negative for spacelike geodesics. For timelike geodesics ($m^2>0$), right-hand side of (\ref{radial}) becomes negative for large
$r$, so that $r$ is bounded and we may extend geodesics
infinitely. For null geodesics, $\dot{r}$ is bounded\footnote{If we allow $r$ to take negative values, then for $E\neq 0$ or $j\neq 0$ the right-hand side becomes oscillatory for large negative $r$, taking positive as well as negative values. Hence both timelike and null geodesics are bounded from below in $r$. For $E=j=0$ the right-hand side is either constant zero or always negative.}. So for any finite values of the affine parameter,
$\dot{r}$ remains finite and so do $\dot{z},\dot{\phi}$ and
$\dot{t}$. Hence the Petrov spacetime is geodesically complete. It will be incomplete if we interpret it as the exterior solution of a rotating cylinder and cut off the region described by the Petrov solution at some value of $r$.
\\
\\By very similar arguments we can show geodesic completeness of the Kaigorodov-Ozsv\'ath solution with line element (\ref{poincare}). In this case, the Lagrangian is
\bequ
\mathcal{L}=\frac{1}{\rho^2}\left(\dot{\rho}^2+\dot{z}^2+\dot{\phi}^2-\dot{t}^2\right)-2\rho\,\dot{z}(\dot{t}+\dot{\phi})+\half\rho^4(\dot{t}+\dot{\phi})^2
\eequ
and the conserved quantities are
\bequ
P\equiv \frac{1}{\rho^2}\dot{z}-\rho(\dot{t}+\dot{\phi}),\;j\equiv
\frac{1}{\rho^2}\dot{\phi}-\rho\dot{z}+\half\rho^4(\dot{t}+\dot{\phi}),\;E\equiv \frac{1}{\rho^2}\dot{t}+\rho\dot{z}-\half\rho^4(\dot{t}+\dot{\phi})
\eequ
so that the radial equation is
\bequ
\left(\frac{d\rho}{d\lambda}\right)^2=-\rho^4(P^2+j^2-E^2)-2\rho^7 P(j+E)-\half\rho^{10}(j+E)^2-\rho^2 m^2.
\label{radial2}
\eequ
The right-hand side becomes negative if $m^2>0$ for both $\rho\rightarrow\infty$ and $\rho\rightarrow 0$, hence all timelike geodesics are bounded in $\rho$. For $m=0$ the right-hand side is either constant zero or becomes negative for large $r$, so null geodesics are bounded from above in $\rho$. For very small $\rho$, $\dot{\rho}$ goes to zero, so as before $\dot{\rho}$ is bounded for null geodesics. This shows geodesic completeness. Note that in the case of anti-de Sitter space (\ref{radial2}) would be 
\bequ
\left(\frac{d\rho}{d\lambda}\right)^2=-\rho^4(P^2+j^2-E^2)-\rho^2 m^2
\eequ
and depending on the magnitudes of $E$, $j$ and $P$ the right-hand side blows up as $\rho\rightarrow\infty$ for some geodesics, which can reach $\rho=\infty$ in finite affine parameter distance.

\subsection{Analysis of Geodesics}
In this subsection we restrict $r$ to be positive for the Petrov solution, in view of the physical requirement (\ref{rhobig}). Outside of the ``horizons'' where $g_{\phi\phi}=g_{tt}=0$ we can write (\ref{radial}) as
\bequ
\left(\frac{dr}{d\lambda}\right)^2=e^{-r}\cos\sqrt{3}r(E-V_{eff}^+(r))(E-V_{eff}^-(r)),
\eequ
where for $j\neq 0$
\bequ
V_{eff}^{\pm}(r)=j\left(\tan\sqrt{3}r\pm\sqrt{\frac{1}{\cos^2\sqrt{3}r}+\frac{e^r}{j^2\cos\sqrt{3}r}(e^{2r}P^2+m^2)}\right)
\eequ
and for $j=0$
\bequ
V_{eff}^{\pm}(r)=\pm\sqrt{\frac{e^r}{\cos\sqrt{3}r}(e^{2r}P^2+m^2)}.
\eequ
The points where $E=V_{eff}^{\pm}(r)$ determine the turning points of
the motion. Near horizons where
the cosine goes to zero, either $V_{eff}^-$ or $V_{eff}^+$ has a
single pole, however the other quantity remains finite as long as
$j\neq 0$. Either a ``particle'' ($E>0$) or an
``antiparticle'' ($E<0$) with some ``angular momentum'' $j$ can reach and cross the horizon. 
\begin{figure}[htp]
\centering
\includegraphics[scale=0.75]{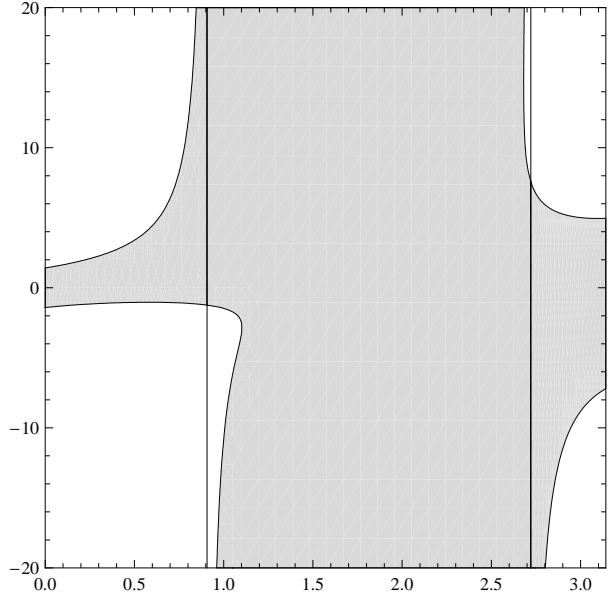}
\caption{{\small A plot of the effective potentials versus $r$ with $j=1$, $P=0$, and $m=1$ for the Petrov spacetime. The gray region is forbidden for this set of parameters.}}
\end{figure}
\\On the horizon, (\ref{radial}) becomes
\bequ
\dot{r}^2=\pm 2e^{-r}Ej-e^{2r}P^2-m^2
\eequ
and hence the horizon can be crossed if
\bequ
E\ge \frac{e^r}{2j}(e^{2r}P^2+m^2)
\eequ
or
\bequ
E\le -\frac{e^r}{2j}(e^{2r}P^2+m^2),
\eequ
which is the value of the finite effective potential branch
on the horizon. Since $E$ cannot be positive and negative at the same
time, no timelike geodesic can reach (and cross) more than one horizon. This statement is independent of the choice of coordinates $t$ and $\phi$ and hence of the position of the horizons. When $j$ is very large, a geodesic can cross one and almost
reach a second horizon before coming back. Again it is clear that timelike geodesics are bounded in $r$.
\\
\\By looking at the radial equation (\ref{radial}) we see that the allowed values of $E$ for geodesic motion lie between $V_{eff}^+$ and $V_{eff}^-$ if the cosine is negative. The first roots of the cosine are at $r\approx 0.907$ and $r\approx 2.721$.
\begin{figure}[htp]
\centering
\includegraphics[scale=0.75]{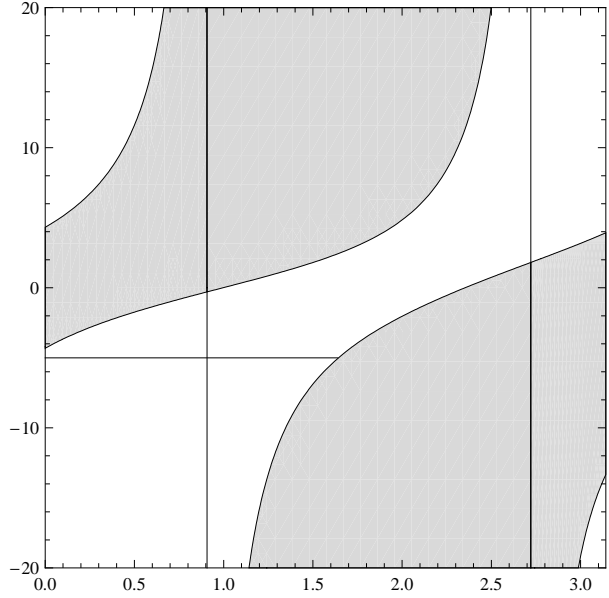}
\caption{{\small Parameters modified to $j=4.2$, $P=0$, and $m=1$, and an ``antiparticle" crossing a horizon of the Petrov spacetime.}}
\end{figure}
\\From the radial equation, we can give the general form for a geodesic:
\bequ
\lambda-\lambda_0=\int\frac{dr}{\sqrt{e^{-r}(E^2\cos\sqrt{3}r-2Ej\sin\sqrt{3}r-j^2\cos\sqrt{3}r)-e^{2r}P^2-m^2}},
\label{hardint}
\eequ
\[z(\lambda)=z_0+\int d\lambda\,e^{2r(\lambda)}P,\;\phi(\lambda)=\phi_0+\int d\lambda\,e^{-r(\lambda)}(j\cos\sqrt{3}r(\lambda)+E\sin\sqrt{3}r(\lambda)),\]
\[t(\lambda)=t_0+\int d\lambda\,e^{-r(\lambda)}(E\cos\sqrt{3}r(\lambda)-j\sin\sqrt{3}r(\lambda)).\]
The integral (\ref{hardint}) can only be solved analytically in special cases. For instance, for $P=0$ and $m=0$ (null geodesics) and $E=j$, $E=0$ or $j=0$ there is a solution in terms of a Gauss hypergeometric function $_2 F_1(a,b;c;z)$. In the extreme case $E=j=0$, requiring $m^2<0$ and hence spacelike geodesics, equation (\ref{hardint}) can be integrated to give (writing $m=iM$)
\bequ
r(\lambda)=\log\left(\left|\frac{\sqrt{M^2(1-\tanh^2(M(\lambda-\lambda_0)))}}{P}\right|\right)
\eequ
In the case of the Kaigorodov-Ozsv\'ath solution with line element (\ref{poincare}) we can write (\ref{radial2}) as \bequ
\left(\frac{d\rho}{d\lambda}\right)^2=\left(\rho^4-\half\rho^{10}\right)(E-V_{eff}^+(\rho))(E-V_{eff}^-(\rho)),
\eequ
where
\bequ
V_{eff}^{\pm}(\rho)=\frac{1}{1-\half\rho^6}\left[\rho^3 P + \half\rho^6 j \pm\sqrt{\half\rho^6 P^2+2\rho^3 P j + P^2 + j^2 +\frac{m^2}{\rho^2}-\half \rho^4 m^2}\right].
\eequ
\begin{figure}[htp]
\centering
\includegraphics[scale=0.75]{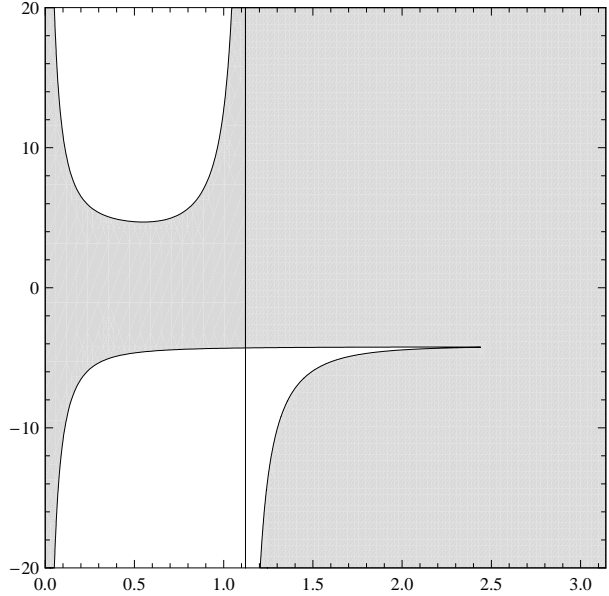}
\caption{{\small Effective potentials versus $\rho$ for the Kaigorodov-Ozsv\'ath spacetime with parameters $j=4.2$, $P=0$, and $m=1$.}}
\end{figure}
\\This expression is ill-defined for $\rho=\sqrt[6]{2}$, when $g_{tt}=0$. However, one branch of the effective potential will remain finite and the situation is similar to the Petrov spacetime, except that there is only a single value of $\rho$ for which $g_{tt}=0$. By similar arguments as before, either positive or negative $E$ is required to cross the ``horizon".
\\
\\The general form of the geodesic is now given by
\bequ
\lambda-\lambda_0=\int\frac{d\rho}{\rho\sqrt{-\rho^2(P^2+j^2-E^2)-2\rho^5 P(j+E)-\half\rho^8(j+E)^2-m^2}},
\label{hardint2}
\eequ
\[z(\lambda)=z_0+\int d\lambda\left(\rho^2(\lambda)P+\rho^5(\lambda)(j+E)\right),\;\phi(\lambda)=\phi_0+\int d\lambda\left(\rho^5(\lambda)P+\half\rho^8(\lambda)(j+E)+\rho^2(\lambda)j\right),\]
\[t(\lambda)=t_0+\int d\lambda\left(-\rho^5(\lambda)P-\half\rho^8(\lambda)(j+E)+\rho^2(\lambda)E\right).\]
Again, the integral (\ref{hardint2}) can only be solved analytically in special cases. One can solve the integral for the case $P=0,\;m=0$ to obtain an implicit definition of $\rho(\lambda)$ which again includes Gauss hypergeometric functions. The case $E=j=0$ gives spacelike geodesics of the form (again $m=iM$)
\bequ
\rho(\lambda)=2\frac{e^{M(\lambda+\lambda_0)}P^2}{e^{2M\lambda_0}+e^{2M\lambda}M^2P^2}.
\eequ

\setcounter{equation}{0}
\section{Holonomy}
The (infinitesimal) {\it holonomy group} of a given spacetime provides another means of classifying solutions of Einstein's equations. Consider an arbitrary point $p$ in a given spacetime, and parallelly transport a tangent vector at $p$ around a closed curve (which is homotopic to the identity, i.e. can be continuously shrunk to a point) through $p$. This defines a linear transformation acting on the tangent space at $p$. The set of all these transformations for different tangent vectors and different curves is a subset of the Lorentz group as parallel transport keeps the norm of a vector constant. One can show that this set is actually a group and hence a subgroup of the Lorentz group.
\\Furthermore, holonomy groups at different points $p,p'$ are isomorphic and hence one can talk about the holonomy group of a given spacetime. By considering the Lie algebra of the holonomy group, one can determine its generators by considering the Riemann tensor and its derivatives. All possible holonomy groups in four dimensions were classified in \cite{schell}.
\\For the Petrov spacetime, we choose to work in the vierbein basis of left-invariant one-forms given by (\ref{1forms})
\[\lambda^1=dr,\;\lambda^2=e^{-r}\,dz\]
\[\lambda^3=e^{\frac{r}{2}}\left(\cos\left(\frac{\sqrt{3}}{2}r\right)\,d\phi-\sin\left(\frac{\sqrt{3}}{2}r\right)\,dt\right),\;\lambda^4=e^{\frac{r}{2}}\left(\sin\left(\frac{\sqrt{3}}{2}r\right)\,d\phi+\cos\left(\frac{\sqrt{3}}{2}r\right)\,dt\right).\]
The Riemann tensor in this basis has non-vanishing components
\bequ
R_{1212}=-1,\;R_{1313}=-R_{1414}=R_{2323}=-R_{2424}=\half,\;R_{1314}=R_{1413}=-R_{2324}=-R_{2423}=\frac{\sqrt{3}}{2}.
\eequ
Following \cite{schell}, we write it as a matrix
\bequ
(R_{AB})=\left( \matrix{-\half &0&0&0&-\frac{\sqrt{3}}{2}&0 \cr 0&-\half &0&-\frac{\sqrt{3}}{2}&0&0 \cr 0&0&1&0&0&0 \cr 0&-\frac{\sqrt{3}}{2}&0&\half&0&0 \cr -\frac{\sqrt{3}}{2}&0&0&0&\half&0 \cr 0&0&0&0&0&-1}\right).
\eequ
As this matrix has full rank, it determines six linearly independent generators of the holonomy group of the Petrov spacetime. Hence the holonomy group must be the Lorentz group $SO(3,1)$.
\\
\\For the Kaigorodov-Ozsv\'ath spacetime we introduce the vierbein basis, in the coordinates of (\ref{lambda}),
\bequ
\omega^1=\sqrt{-\frac{3}{\Lambda}}dr,\quad\omega^2=e^{-r}dz-\sqrt{2}e^{2r}d\phi,\quad\omega^3=e^{-4r}dt,\quad\omega^4=e^{2r}d\phi-e^{-4r}dt,
\eequ
so that indeed $\eta_{\mu\nu}\omega^{\mu}\otimes\omega^{\nu}=g_{ab}dx^a\otimes dx^b$. In this frame the Riemann tensor takes the form, written as a matrix,
\bequ
(R_{AB})=\left(-\frac{\Lambda}{3}\right)\cdot\left( \matrix{-\half &0&0&0&\frac{3}{2}&\frac{3}{\sqrt{2}} \cr 0&\frac{5}{2} &\frac{3}{\sqrt{2}}&\frac{3}{2}&0&0 \cr 0&\frac{3}{\sqrt{2}}&1&\frac{3}{\sqrt{2}}&0&0 \cr 0&\frac{3}{2}&\frac{3}{\sqrt{2}}&\half&0&0 \cr\frac{3}{2}&0&0&0&-\frac{5}{2}&-\frac{3}{\sqrt{2}}\cr \frac{3}{\sqrt{2}}&0&0&0&-\frac{3}{\sqrt{2}}&-1}\right).
\eequ
The trace of the top-left ($3\times 3$) block matrix is $-\Lambda$, in agreement with \cite{schell}. This matrix also has full rank, therefore the Kaigorodov-Oszv\'ath spacetime also has holonomy group $SO(3,1)$.

\section{Conclusions}
We have given a matrix representation of the Petrov spacetime, identified with its four-dimensional isometry group, and calculated a basis of left-invariant one-forms from which invariant actions could be constructed as a starting point of non-linear realisations. The Killing algebra admits a two-parameter family of linear deformations which lead to three distinctive cases. One of these gives the isometry group of a different Einstein manifold, the Kaigorodov-Ozsv\'ath solution, and we could construct the metric on this manifold from the basis of left-invariant one-forms. A solution of this type is known to exist only for negative cosmological constant.
\\When discussing non-linear realisations from the given bases of left-invariant one-forms, we showed that by eliminating the non-dynamical variable from a general first-order action one obtains a Lagrangian of Finsler type in both cases.
\\A physical interpretation of the Petrov spacetime is given by its identification with an exterior solution of an infinite rigidly rotating dust cylinder of van Stockum type. In the context of causality violation we have shown that one cannot create CTCs by spinning up a cylinder beyond its critical angular velocity by shooting in particles on timelike or null curves.
\\We have shown that both the Petrov and the Kaigorodov-Ozsv\'ath spacetime are geodesically complete and made general statements about geodesics on these manifolds using the radial equation and effective potentials. Explicit solutions of the appearing integrals could only be given for very special cases.
\\Both spaces were shown to have maximal holonomy group $SO(3,1)$. 
\\The simple examples of homogeneous space-times presented in this paper show how non-linear realisations of spacetime symmetries may be implemented in general and how deformations of the isometry group can relate different physical situations to each other. It is hoped that these considerations will prove to be helpful in more physically interesting setups.

\section*{Acknowledgements}
SG would like to thank for financial support from EPSRC and Trinity College, Cambridge, and to thank Claude Warnick for helping with the figures. GWG would like to thank Joaquim Gomis for a number of valuable comments and suggestions.
\\We wish to thank one of the referees for pointing out a few minor errors in an earlier version of this paper.

\end{document}